\documentstyle[12pt]{article}

\newcommand{\be}{\begin{eqnarray} }
\newcommand{\ee}{\end{eqnarray} }

\begin{document}
\begin{center}
{\large\bf Discussion on \\ "Dynamical Chern-Simons term generation at
finite density"\\ and "Chern-Simons term at finite density"}
\vskip 1cm

{\large A.N.Sissakian,}\\
{\it Bogolubov Theoretical Laboratory,
Joint Institute for Nuclear Research,\\
Dubna, Moscow region 141980, Russia}\\
\vskip 5mm

{\large O.Yu.Shevchenko\footnote{E-mail address: shevch@nusun.jinr.ru}
and S.B.Solganik\footnote{E-mail address: solganik@thsun1.jinr.ru} }\\
{\it Laboratory of Nuclear Problems, Joint Institute for Nuclear Research,\\
Dubna, Moscow region 141980, Russia}
\end{center}

\vskip 1cm
\abstract
We discuss the comment by V.Zeitlin on our recent papers
concerning Chern-Simons term generation at finite density.

\vskip 1cm

The point of the V.Zeitlin's comment \cite{ze} is contradiction of our
result \cite{my} with
the results obtained in papers \cite{tolpa} for $QED_{2+1}$.
However, it isn't quite correct statement.
As in  papers \cite{tolpa} so as in our articles \cite{my}  mathematics is
consistent. However, the investigated objects are essentially
different.
In  \cite{tolpa} it  was calculated not parity
violating covariant form in action -- Chern-Simons (CS) topological
term, but the charge density which
includes  as parity odd  as well parity even parts
\footnote{When we, for abbreviation,
speak about parity invariance properties of local objects we, certainly,
have in mind  symmetries of the corresponding action parts.}.
The parity odd part (we are studding) leads
to CS term and mass generation of the gauge field
in effective action. The parity even part reads
\be
J^{0}_{\rm even}=\frac{|eB|}{2\pi}\left( {\rm Int}\left[
\frac{\mu^{2}-m^{2}}{2|eB|}
\right]+\frac{1}{2}\right) \theta (\mu -|m|)
\ee
(which is parity invariant because under parity $B\rightarrow -B$).
It is clear that this parity even term
does  contribute
neither to the parity anomaly nor to the mass of the
gauge field.
In our articles we are interested in
parity odd covariant topological CS term in action, changing on
winding number under
gauge transformation and leading to the  mass generation of a gauge field.

Moreover, in \cite{tolpa}  calculations are done  in the pure
magnetic background and scalar magnetic field occurs
in the argument's denominator of the cumbersome function.
So, the parity even term seems to be
''noncovariantizable'', i.e. it can't be converted in covariant
form in effective action.

In our  papers \cite{my}  CS term was calculated at finite density
as in abelian  so as in nonabelian gauge theory.
Besides, in \cite{my}  the calculations
were performed in apparently covariant form for arbitrary background
gauge field.
Calculations were done as in 3--dimensional
theory so as in 5--dimensional one.
The result was generalized to any odd dimensional gauge theory.
It was shown that $\mu$--dependent CS term coefficient
reveals the amazing property of universality. Namely, it does depend on
neither dimension of the theory nor abelian or nonabelian gauge
theory is studied.

In our papers \cite{my} it was shown that in any odd dimension CS term coefficient
has the same form
\begin{equation}
\label{kon}
S^{{\rm C.S}}_{eff}=\frac{m}{|m|}\theta (m^2 -\mu^2 ) \pi W[A] ,
\end{equation}
where $W[A]$ is the CS term in any odd dimension.
Since  only the lowest orders
of perturbative series contribute to CS term at finite density
(the same situation
is well-known at zero density), the result obtained by using
formally perturbative technique appears to be nonperturbative.
Thus, the $\mu$--dependent CS term coefficient
reveals the amazing property of universality.
Namely, it does depend on
neither dimension of the theory nor abelian or nonabelian gauge
theory is studied.

The arbitrariness of $\mu$ gives us the possibility
to see Chern--Simons coefficient behavior at any masses.
It is very interesting that  $\mu^2 = m^2$ is the
crucial point for Chern--Simons.
Indeed, it is clearly seen from (\ref{kon}) that when $\mu^2 < m^2$
$\mu$--influence disappears and we get the usual Chern-Simons term
$I^{{\rm C.S}}_{eff}= \pi W[A].$
On the other hand when $\mu^2 > m^2$
the situation is absolutely different.
One can see that here the Chern-Simons term
disappears because of non--zero density of background fermions.
We'd like to emphasize the
important massless case $m=0$ considered in \cite{ni1}.
Then even negligible density,
which always take place in any
physical processes, leads to vanishing of the parity anomaly.

In conclusion, let us stress again that we nowhere have used
any restrictions on $\mu$.
Thus we not only confirm result
in \cite{ni1} for Chern--Simons in $QED_{3}$ at small density,
but also expand it
on arbitrary $\mu$, nonabelian case and arbitrary odd dimension.

\end{document}